\documentclass[aps,pra,reprint,superscriptaddress,showpacs]{revtex4-1}

\usepackage{latexsym}
\usepackage{graphics}
\usepackage{amsmath}
\usepackage{color}
\usepackage{graphicx}

\usepackage{array}

\begin{document}

\title[Reconciling simulated melting and ground-state properties of metals]
{Reconciling simulated melting and ground-state properties of metals with
a modified embedded-atom method potential}

\author{Gennady Sushko}
\affiliation{MBN Research Center, Altenh\"oferallee 3, 60438 Frankfurt am Main, Germany}

\author{Alexey Verkhovtsev}
\email[]{verkhovtsev@fias.uni-frankfurt.de}
\affiliation{MBN Research Center, Altenh\"oferallee 3, 60438 Frankfurt am Main, Germany}
\affiliation{Frankfurt Institute for Advanced Studies, Ruth-Moufang-Str. 1, 60438 Frankfurt am Main, Germany}

\author{Christian Kexel}
\affiliation{MBN Research Center, Altenh\"oferallee 3, 60438 Frankfurt am Main, Germany}
\affiliation{Department of Physics, Goethe-Universit\"at, Max-von-Laue-Str. 1, 60438 Frankfurt am Main, Germany}

\author{Andrei V. Korol}
\affiliation{MBN Research Center, Altenh\"oferallee 3, 60438 Frankfurt am Main, Germany}
\affiliation{Department of Physics, St. Petersburg State Maritime Technical University,
Leninskii prospekt 101, 198262 St. Petersburg, Russia}

\author{Stefan Schramm}
\affiliation{Frankfurt Institute for Advanced Studies, Ruth-Moufang-Str. 1, 60438 Frankfurt am Main, Germany}
\affiliation{Department of Physics, Goethe-Universit\"at, Max-von-Laue-Str. 1, 60438 Frankfurt am Main, Germany}

\author{Andrey V. Solov'yov}
\altaffiliation{On leave from A.F. Ioffe Physical-Technical Institute,
Politekhnicheskaya ul. 26, 194021 St. Petersburg, Russia}
\affiliation{MBN Research Center, Altenh\"oferallee 3, 60438 Frankfurt am Main, Germany}

%\date{\today}

\begin{abstract}
We propose a modification of the embedded-atom method-type potential aiming at
reconciling simulated melting and ground-state properties of metals by means of
classical molecular dynamics.
Considering titanium, magnesium, gold, and platinum as case studies,
we demonstrate that simulations performed with the modified force field
yield quantitatively correctly both the melting temperature of the metals
and their ground-state properties.
It is shown that the accounting for the long-range interatomic interactions
noticeably affect the melting point assessment.
The introduced modification weakens the interaction at interatomic distances
exceeding the equilibrium one by a characteristic vibration amplitude defined
by the Lindemann criterion, thus allowing for the correct simulation of melting,
while keeping its behavior in the vicinity of the ground state minimum.
The modification of the many-body potential has a general nature and can be
applicable to metals with different characteristics of the electron structure
as well as for many different molecular and solid state systems experiencing
phase transitions.
\end{abstract}

\maketitle

%%%%%%%%%%%%%%%%%%%%%%%%%%%%%%%%%%%%%%%%%%%%%%%%%%%%%%%%%%%%%%%%%%%%%
%%%%%%%%%%%%%%%%%%%%%%%%%%%%%%%%%%%%%%%%%%%%%%%%%%%%%%%%%%%%%%%%%%%%%
\section{Introduction}

The melting of crystals and crystallization of liquids are of great scientific
and technological significance.
The solid-liquid phase boundary represents an important part of the phase diagram
and is widely explored in material science, high-pressure physics, astrophysics,
and geophysical sciences \cite{Bang_2010_AdvMater.22.1039, Yoo_1993_PRL.70.3931,
Belonoshko_2008_PRL.100.135701}.
Along with experimental methods of studying phase transitions,
classical molecular dynamics (MD) simulations \cite{Rapaport_Art_of_MD, Leach_MolModel}
represent a powerful tool which have an eminent research potential.
It can provide insights into nanoscale structural features and
thermo-mechanical properties of the system under study by means of advanced
computer simulations \cite{MBN_Explorer1}.
Provided that interatomic potentials, which are used to model interactions
in a system, correctly describe different system properties, classical MD
simulations may become a low-cost alternative to experimental studies
and allow one to reach the system sizes and time scales that are inaccessible
by {\it ab initio} methods
\cite{Schulten_2013_Nature.497.643, Vashishta_2006_JPhysChemB.110.3727,
Verkhovtsev_2013_ComputMaterSci.76.20, Yakubovich_2013_PhysRevB.88.035438}.

Despite the abundance of interatomic potentials for modeling metallic, organic,
and biomolecular systems, and complex systems composed of such constituents
\cite{MBN_Explorer1, Rafii-Tabar_potentials, MacKerell_1998_CHARMM, Monticelli_potentials},
the overwhelming majority of these functions are capable of reproducing
only ground-state properties of a system.
While matching the results of {\it ab initio} calculations of ground-state
parameters, these force fields poorly describe highly-excited vibrational
states when the system under study is far from the potential energy minimum,
that is the case when a phase transition occurs in the system
\cite{Lindemann_1910_ZPhys.11.609, AdvChemPhys_137_2008}.
The proper quantitative description of phase transitions in general and
the melting process in particular by means of MD simulations is a major
scientific challenge that concerns metal materials
\cite{Los_2010_PhysRevB.81.064112, Kexel_2015_EPJB.88.221},
as well as inorganic \cite{Hussien_2010_EPJD.57.207}, and biomolecular systems,
like proteins \cite{Schulten_2010_NatPhys.6.751}
or water \cite{Vega_2011_PCCP.13.19663, Shvab_2012_PhysRevE.85.051509}.

This paper aims at formulating a recipe for constructing an interatomic potential
that is able to correctly reproduce both the melting temperature and the
ground-state properties of metal systems by means of classical MD simulations.
To achieve this goal, we propose a modification of the widely utilized
embedded-atom method (EAM)-type potential
\cite{Daw_1984_PhysRevB.29.6443, Daw_1993_MaterSciRep.9.251}
and demonstrate its applicability to different metal systems.
Our analysis has revealed that interatomic interactions at distances,
exceeding the equilibrium distance by a characteristic vibration amplitude
defined by the Lindemann melting criterion
\cite{Lindemann_1910_ZPhys.11.609, Nelson_Defects_and_Geometry},
significantly affect the correctness of simulations.
In order to reproduce accurately the value of the melting point,
these interactions should be corrected as they are overestimated
by conventional EAM-type potentials.
The modified force field weakens the interatomic interactions at distances beyond
the equilibrium point, thus yielding the correct value of melting temperature.

\section{Computational details}

In the EAM approach, the total energy of a metal system is expressed via
the energy $F_i$, obtained by embedding an atom $i$ into the local electron density
$\bar{\rho}_i$ provided by the remaining atoms of the system, and that of the short-range
electrostatic interaction between atoms $i$ and $j$ separated by a distance $r_{ij}$:
\begin{equation}
U(r_{ij})
= \sum_{i=1}^N \left[ F_i (\bar{\rho}_i) + \frac12 \sum_{j \ne i} \phi(r_{ij}) \right] \ .
\label{EAM}
\end{equation}

\noindent
Different many-body potentials \cite{Finnis-Sinclair, Sutton-Chen, Gupta, TB-SMA_2},
that have such a general form,
are capable of describing geometrical, mechanical, and energetic properties
(e.g., cohesive energy, lattice parameters, and elastic constants),
but can rarely reproduce the experimentally measured melting temperature.
An illustrative case study is bulk titanium, whose melting temperature,
as calculated using the many-body potentials which account for the interaction
of a given atom with several surrounding atomic layers
\cite{Kim_2006_PhysRevB.74.014101, Baskes_1994_ModelSimulMaterSciEng.2.147},
differ from experimental values by several hundred degrees
\cite{Kim_2006_PhysRevB.74.014101, Sushko_2014_JPhysChemA_diff}.
A similar order of discrepancy was observed also for other systems,
such as gold and silicon, modeled using the original and more elaborated EAM
potentials \cite{Lewis_1997_PRB.56.2248, Ryu_2009_ModelSimulMaterSciEng.17.075008}.
Thus, it is essential to amend the existing force fields, so that they can reproduce
correctly properties of both the ground- and finite-temperature
states of metal systems.

In this work, we propose such a modification.
The new potential $U_{\rm mod}(r_{ij})$ should satisfy the principal condition that
the curvature of the modified potential energy profile in the vicinity of the
equilibrium point must coincide with that of the original potential.
This condition is set to reproduce, with the new potential, ground-state
properties which are governed by the behavior of the potential energy curve in the
vicinity of the equilibrium point.

To properly simulate melting of metal systems, we introduce a modification
that satisfies the above-defined condition.
As an illustration, we add a linear function to the existing formula,
so that the modified expression for the potential energy of the system
reads as:
\begin{equation}
U_{\rm mod}(r_{ij}) = U(r_{ij}) + B \, r_{ij} + C \ ,
\label{Finnis_Sinclair_pot_mod}
\end{equation}

\noindent
where $U(r_{ij})$ is the original EAM-type potential (\ref{EAM}),
and $B$ and $C$ are adjustable parameters.
As shown below, the term $B \, r_{ij}$ makes the resulting potential steeper
(less attractive) at interatomic distances exceeding the equilibrium point but,
at the same time, it also slightly changes the depth of the potential well at
the equilibrium point.
The constant term $C$ is thus introduced to discard the latter effect.
As a case study, we used the exact form of the potential $U(r_{ij})$ which
is based on the second-moment approximation of the tight-binding model
\cite{Rafii-Tabar_potentials}.
According to this scheme \cite{TB-SMA, TB-SMA_2},
the attractive many-body part of the potential is related to band energy and
is expressed as a square-root of electron density,
$F_i(\bar{\rho}_i) \sim \sqrt{\bar{\rho}_i}$.
Both the attractive and repulsive terms are introduced in this approach
in the exponential form \cite{TB-SMA, Tomanek_1985_PhysRevB.32.5051}
which is commonly referred in the literature as to the Gupta potential \cite{Gupta}.
The total potential energy of a system of $N$ atoms located at positions ${\bf r}_i$
reads as:
\begin{eqnarray}
U &=& \sum_{i=1}^N \left\{ \sum_{j \ne i} A\,
\exp{ \left[ -p \, \left( \frac{r_{ij}}{d} - 1 \right) \right]}  \right. \nonumber \\
&-&
\left.
\sqrt{ \sum_{j \ne i} \xi^2\,
\exp{ \left[ -2q \, \left( \frac{r_{ij}}{d} - 1 \right) \right] } } \, \right\} \ .
\label{Finnis_Sinclair_pot}
\end{eqnarray}

\noindent
Here,
$d$ is the first-neighbor distance,
$\xi$ is an effective overlap integral between electronic orbitals
of neighboring atoms,
$q$ and $p$ control the decay of the exponential functions and are related
to bulk elastic constants \cite{Tomanek_1985_PhysRevB.32.5051}.
We note that the introduced modification~(\ref{Finnis_Sinclair_pot_mod})
is spiritually similar to the well-known Dzugutov pairwise
potential~\cite{Dzugutov_1992_PRA.46.2984} which was developed to model
glass-forming liquid metals.
The Dzugutov potential is constructed to suppress crystallization common to
most monatomic systems by the introduction of a repulsive term representing
the Coulomb interactions that are present in a liquid metal.
The similar idea of increasing repulsion at large interatomic distances
for modeling metals in highly-vibrational states far from the potential energy minimum
is pursued with the introduced linear modification.

The impact of the modified potential was investigated by analyzing thermal,
geometrical, and energetic properties of nanoscale samples composed of four
representative metals, namely titanium, magnesium, gold, and platinum.
A generality of the introduced correction is emphasized by considering metals
with different characteristics of the electron structure, namely
(i) $s,p-$bonding (Mg),
(ii) transition metal with less than half-filled $d$ band (Ti),
(iii) transition metal with almost filled $d$ band (Pt), and
(iv) noble metal (Au).

\begin{table*}[htb!]
\centering
\caption{Utilized parameters of the original (\ref{Finnis_Sinclair_pot})
and the modified (\ref{Finnis_Sinclair_pot_mod}) EAM-type potentials
describing the interactions in titanium, magnesium, gold, and platinum.}
\begin{tabular}{p{1.0cm}p{1.0cm}p{1.3cm}p{1.0cm}p{1.3cm}p{1.0cm}p{1.0cm}p{1.6cm}p{1.6cm}p{1.0cm}}
\hline
     & $d$~(\AA) & $A$~(eV) & $p$ & $\xi$~(eV) & $q$ & Ref. & $B$~(eV/\AA) & $C$ (eV) & $r_c$~(\AA) \\
\hline
 Ti & 2.95 & 0.153 & 9.25  & 1.88 & 2.51 & \cite{Lai_2000_JPhysCondMatter.12.L53} &  0.0114  & $-0.060$  &  7.0  \\
 Mg & 3.21 & 0.029 & 12.82 & 0.50 & 2.26 & \cite{TB-SMA}                          &  0.0061  & $-0.032$  &  7.0  \\
 Au & 2.88 & 0.206 & 10.23 & 1.79 & 4.04 & \cite{TB-SMA}                          &  0.0065  & $-0.034$  &  6.65 \\
 Pt & 2.78 & 0.297 & 10.61 & 2.70 & 4.00 & \cite{TB-SMA}                          &  0.0064  & $-0.031$  &  6.6  \\
\hline
\end{tabular}
\label{Table_FF_parameters}
\end{table*}

We considered finite-size spherical nanoparticles with radii
from 1 to 7~nm, cut from ideal hexagonal close-packed (hcp) (in the case of Ti and Mg)
or face-centered cubic (fcc) (for Au and Pt) crystals.
The nanoparticles were composed of approximately 300 to 80\,000 atoms.
The crystalline structures were constructed and optimized, and the
MD simulations were carried out
using the MBN Explorer software package~\cite{MBN_Explorer1}.
Energy minimization was performed using the velocity-quenching algorithm.
The MD simulations of the nanoparticle heating/melting were performed
without boundary conditions in the NVT canonical ensemble.
The temperature $T$ was controlled by a Langevin thermostat with a damping
coefficient of 1~ps$^{-1}$.
The nanoparticles were heated up with a constant rate of 0.5~K/ps.
The time integration of the equations of motion was done using the velocity-Verlet
algorithm \cite{Frenkel_2002_MD} with an integration time step of 5~fs.
In all the calculations, the interatomic interactions were truncated at the cutoff
radius $r_c$ lying in the range between 6.6 and 7~\AA, depending on the system.
Parameter $B$ was derived independently for each considered metal so that the
extrapolated bulk melting point corresponds to the reference value.
The parameter $C$ was then tuned to reproduce the reference value of cohesive energy.
Parameters of the potential~(\ref{Finnis_Sinclair_pot}) and
the correction~(\ref{Finnis_Sinclair_pot_mod}) utilized in this work
are summarized in Table~\ref{Table_FF_parameters}.

In the proposed modification, the linear term $B r_{ij} + C$
is responsible for a monotonic increase of the potential at large distances.
In this case the cutoff distance is set to the value at which the modified
potential (\ref{Finnis_Sinclair_pot_mod}) is equal to zero.
The parametrization of the original EAM-type potential for titanium, given in
Table~\ref{Table_FF_parameters}, was obtained in Ref.~\cite{Lai_2000_JPhysCondMatter.12.L53}
with the cutoff distance of 4.2~\AA~as another adjustable parameter.
The other three metals are described in this work with the parametrization by Cleri
and Rosato~\cite{TB-SMA} where the summation in the EAM-type potential was
''.. extended up to the fifth neighbors for cubic structures''.
The analysis of radial distribution function for gold and platinum demonstrates that
the fifth neighbors in these metals are located at the distance 6.45 and 6.15~\AA~from
the given atom, respectively.
These values are slightly smaller than the cutoff values which we have used in the
simulations, see Table~\ref{Table_FF_parameters}.
In reference~\cite{TB-SMA}, hcp metals were described ''... with cutoff values ranging
between $\sqrt{11/3} \, d$  and $\sqrt{5} \, d$'' where $d$ is the first-neighbor distance.
The original cutoff for titanium, as formulated in Ref.~\cite{TB-SMA}, thus lies
in the range from 5.65 to 6.60~\AA~which is smaller than the cutoff used in our simulations.
Similarly, the original cutoff for magnesium lies in the range between 6.15 and 7.2~\AA~and
corresponds to the value of $r_c = 7$~\AA~which we have adopted in the simulations.

In theory, a cutoff distance should be set on the grounds that forces at larger
interatomic distances are negligibly small, so that distant interactions are
infinitesimally weak and could be excluded from consideration.
However, there is no unique way to define the cutoff distance in a general case;
thus, the value utilized in every simulation defines its accuracy and corresponding
computational costs.
We found that the ground-state properties like lattice constants are nicely described
even with small values of cutoff
(e.g., $r_c = 4.2$~\AA~for titanium~\cite{Lai_2000_JPhysCondMatter.12.L53}), while
the estimated value of melting temperature turned out to be cutoff-dependent.
Our analysis has demonstrated that an explicit account of very distant interatomic
interactions when using the original EAM-type potential (\ref{Finnis_Sinclair_pot})
does not allow for a proper quantitative description of melting, and the potential
modification is required to bring the calculated melting temperature closer to the
experimental values.
In the performed simulations, the cutoff distance is set to the value at which the
modified potential is equal to zero, thereby significantly reducing cutoff effects
in the modified potential compared to the original approach.

\section{Results and discussion}

To quantify the effect due to the potential modification, we have analyzed first
the ground-state geometrical and energetic properties of the samples,
namely lattice parameters and cohesive energy (see table~\ref{Table_sim}).
The quantity $E_{\rm coh}$ represents the cohesive energy per atom of an
infinitely large ideal crystal, which was obtained by extrapolating the
binding energies of Ti$_N$, Mg$_N$, Au$_N$ and Pt$_N$ ($N \approx 300 - 80\,000$)
nanoparticles to the $N \to \infty$ limit.
Table~\ref{Table_sim} demonstrates that, similar to the case of the original potential
(the column labeled as "$U_{\rm orig}$"),
the modification of the potential (the column "$U_{\rm mod}$")
allows one to reproduce experimental values \cite{Kittel} with a relative discrepancy
of less than~2\%.

\begin{table}
\centering
\caption{Comparison of ground-state parameters of the modeled crystals for
different specifications of the potential.}
\begin{tabular}{p{1.5cm}p{0.9cm}p{0.9cm}p{1.3cm}p{0.9cm}p{0.9cm}p{0.9cm}}
\hline
                 &  $U_{\rm orig}$ & $U_{\rm mod}$ & exp. &  $U_{\rm orig}$  & $U_{\rm mod}$  &  exp. \\
\hline
                    & \multicolumn{3}{c}{Ti}  & \multicolumn{3}{c}{Mg}  \\
\hline
$a$, \AA            &  2.92  &  2.92  &  2.95  &  3.20  &  3.20  & 3.21  \\
$c$, \AA            &  4.76  &  4.76  &  4.68  &  5.22  &  5.22  & 5.21  \\
$E_{\rm coh}$, eV   &  5.04  &  4.85  &  4.87  &  1.52  &  1.49  & 1.52  \\
\hline
\hline
                    & \multicolumn{3}{c}{Au}  & \multicolumn{3}{c}{Pt} \\
\hline
$a$, \AA            &  4.07  &  4.05  &  4.08  &  3.93  &  3.91  & 3.92  \\
$E_{\rm coh}$, eV   &  3.78  &  3.80  &  3.78  &  5.85  &  5.84  & 5.85 \\
\hline
\end{tabular}
\label{Table_sim}
\end{table}

%%%%%%%%%%%%%%%%%%%%%%%%%%%%%%%%%%%%%%%%%%
%\subsection*{Vacancy formation energy}

As another benchmark of the modified potential, we have also analyzed vacancy
formation energy. This quantity is given by
\begin{equation}
E_{\rm vf} = E_{N-1} - \frac{N - 1}{N} \, E_N  \ .
\end{equation}
where $E_N$ and $E_{N-1}$ are the energies of a perfect crystal and a vacancy-formed
structure after relaxation, and $N$ is the number of atoms in the simulation box.
To calculate $E_{\rm vf}$, the following procedure was adopted.
A perfect crystal was created, which spans at least three cutoff distances in each direction.
The crystal comprising $N$ atoms was relaxed using periodic boundary conditions.
Then, one atom was removed from the crystal.
The crystal now comprising $N - 1$ atoms was relaxed again using periodic boundary conditions.
To check the consistency of the results, we have analyzed the samples of different size,
containing from about 500 up to 2048 atoms.

Table~\ref{Table_VFE} presents the vacancy formation energy calculated with the original
(the column ''$U_{\rm orig}$'') and the modified (the column ''$U_{\rm mod}$'') potentials.
The calculated values are compared with available experimental data and the results
of earlier DFT and classical calculations.
This analysis demonstrates that the numbers obtained with the original EAM-type
potential~(\ref{Finnis_Sinclair_pot}) and its modification (\ref{Finnis_Sinclair_pot_mod})
are consistent with one another and agree in general with the existing experimental and
theoretical values.

\begin{table*}[htb!]
\centering
\caption{
Vacancy formation energy (in eV) calculated with the original ($U_{\rm orig}$)
and the modified ($U_{\rm mod}$) potential.
The calculated values are compared with available experimental data and the results of
earlier calculations.
The experimental methods comprise positron annihilation (PA), thermal expansion (TE) and
quenching (Q) measurements.
Earlier theoretical calculations performed by means of density functional theory are labeled
as DFT, and LDA/GGA denote the local density or generalized gradient approximations.
EAM denotes the classical MD simulations performed using an EAM-type potential.
}
\begin{tabular}{p{0.8cm}p{1.0cm}p{1.0cm}p{3.5cm}p{4.0cm}}
\hline
       & \multicolumn{2}{c}{this work}  &    &  \\ \cline{2-3}
       &  $U_{\rm orig}$   & $U_{\rm mod}$ &  exp. data  & calculations   \\
\hline
  Ti  &  1.56  &  1.52  & 1.55~\cite{Shestopal_1966_FTT.7.3461}
      & 1.56 (EAM)~\cite{Lai_2000_JPhysCondMatter.12.L53}   \\
      &        &        & $1.27 \pm 0.05$ (PA)~\cite{Hashimoto_1984_JPhysF.14.L215}
      & 2.14 (DFT-LDA)~\cite{LeBacq_1999_PhysRevB.59.8508}  \\
      &        &        &
      & 1.97 (DFT-GGA)~\cite{Raji_2009_PhilosMag.89.1629}   \vspace{0.15cm} \\
  Mg  &  0.60  &  0.62  & $0.58 \pm 0.01$ (TE)~\cite{Janot_1970_PhysRevB.2.3088}
      & 0.88 (EAM)~\cite{Johansen_2009_CompMaterSci.47.121}      \\
      &        &        & $0.79 \pm 0.03$ (Q)~\cite{Tzanetakis_1976_PhysStatSolB.75.433}
      & $0.77 - 0.80$ (DFT-GGA)~\cite{Jund_2013_JPhysCondMat.25.035403} \vspace{0.1cm} \\
  Au  &  0.61  &  0.64  & $0.62 - 0.67$ (TE)~\cite{Jongenburger_1957_PhysRev.106.66}
      & 0.60 (EAM)~\cite{TB-SMA_2}      \\
      &        &        & $0.70 - 1.10$ (Q)~\cite{Jongenburger_1957_PhysRev.106.66}
      & 0.75 (EAM)~\cite{TB-SMA}     \vspace{0.15cm} \\
  Pt  &  1.16  &  1.14  &  $1.35 \pm 0.09$ (PA)~\cite{Schaefer_1987_PhysStatSol.102.47}
      & 1.28 (EAM)~\cite{TB-SMA_2}         \\
      &        &        &
      & 1.15 (DFT-LDA)~\cite{Mattsson_2002_PhysRevB.66.214110}  \\
      &        &        &
      & 1.18 (DFT-GGA)~\cite{Mattsson_2002_PhysRevB.66.214110}  \\
\hline
\end{tabular}
\label{Table_VFE}
\end{table*}

%%%%%%%%%%%%%%%%%%%%%%%%%%%%%%%%%%%%%%%%%%

\begin{figure}[ht]
\centering
\includegraphics[width=0.42\textwidth,clip]{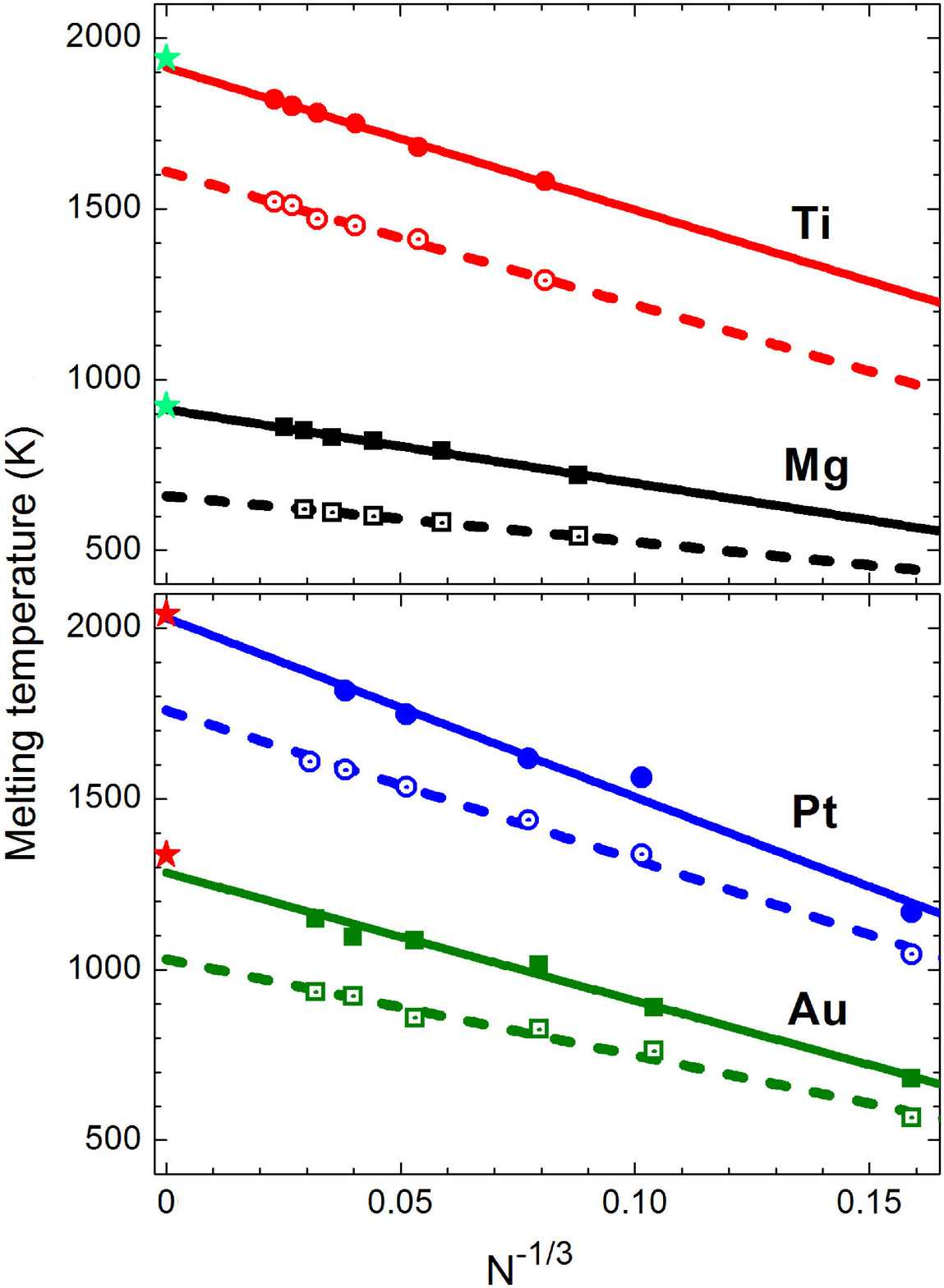}
\caption{
Melting temperature of spherical Ti$_N$, Mg$_N$, Au$_N$, and Pt$_N$ nanoparticles
calculated by means of the original, $U$ (open symbols), and modified, $U_{\rm mod}$
(closed symbols), potential.
Lines represent the linear extrapolation of the calculated numbers to the bulk
($N \to \infty$) limit.
Experimental values of melting temperature are shown by stars.
}
\label{fg:melting}
\end{figure}

Melting temperature of the finite-size nanoparticles was estimated
from analyzing the temperature dependence of the heat capacity,
$C_V = \left( \partial E / \partial T \right)_V$, defined as a
derivative of the internal energy of the system with respect to temperature.
A sharp maximum of $C_V$ was attributed to the nanoparticle melting.
The bulk melting temperature was estimated by extrapolating the obtained
values to the $N \to \infty$ limit according to the Pawlow law
\cite{Pawlow_1909_ZPhysChem.65, Qi_2001_JChemPhys.115.385,
Yakubovich_2013_PhysRevB.88.035438}.
It describes the dependence of the melting temperature of spherical particles
on the number of atoms they are composed of as
$T_{m} = T_{m}^{\rm bulk} - \alpha N^{-1/3}$, where $T_{m}^{\rm bulk}$ is the
melting temperature of a bulk material and $\alpha$ is the factor of proportionality.
Thus evaluated values of melting temperature are summarized in Figure~\ref{fg:melting}
and Table~\ref{Table_Tmelt} for all the studied metals.

In Figure~\ref{fg:melting}, symbols illustrate the results of the simulations
for the finite-size nanoparticles.
The estimated values of the bulk melting temperature obtained with the use of the
original potential (\ref{Finnis_Sinclair_pot}) (open symbols) lead to a significant
deviation of about 300~K from the experimental values which are marked by stars.
The situation changes drastically when introducing the linear correction to the original
potential.
Figure~\ref{fg:melting} illustrates that the use of the modified force field (closed symbols)
leads to a much better correspondence of the bulk-limit extrapolations with the experimental
values for all studied metals.
The extrapolation procedure yields the values of the melting temperature presented in
Table~\ref{Table_Tmelt} are in good agreement with the reference values with the relative
discrepancy of a few ($1-4\%$) percent.

\begin{table}
\centering
\caption{Melting temperature $T_{m}^{\rm bulk}$ (in kelvin) of different metals
which is evaluated on the basis of the performed MD simulations.}
\begin{tabular}{p{1.0cm}p{1.2cm}p{1.2cm}p{1.2cm}}
\hline
         &   $U$  & $U_{\rm mod}$  &  exp. \\
\hline
% old data:
%Mg       &    658  &    913  &  923   \\
%Au       &    1044 &    1344 &  1337  \\
%Ti       &    1610 &    1915 &  1941  \\
%Pt       &    1740 &    2060 &  2041  \\
%
Mg       &    658  &    913  &  923   \\
Au       &    1030 &    1284 &  1337  \\
Ti       &    1610 &    1915 &  1941  \\
Pt       &    1759 &    2030 &  2041  \\
\hline
\end{tabular}
\label{Table_Tmelt}
\end{table}

In order to shed light on the physical effects which are behind the above-described improvement,
we have analyzed melting of the studied metal systems in terms of the Lindemann criterion
\cite{Lindemann_1910_ZPhys.11.609}.
It states that melting occurs because of vibrational instability, i.e. a crystalline structure
melts when the average amplitude of thermal vibrations of atoms is relatively high compared
to interatomic distances.
This condition can be expressed as
$\langle \left( \delta u \right)^2 \rangle^{1/2} > \delta_L d$,
where $\delta u$ is the atomic displacement,
$\delta_L$ is the Lindemann parameter typically equal to $0.10 - 0.15$,
and $d$ is the interatomic distance \cite{AdvChemPhys_137_2008}.

\begin{figure*}[ht]
\centering
\includegraphics[width=0.92\textwidth,clip]{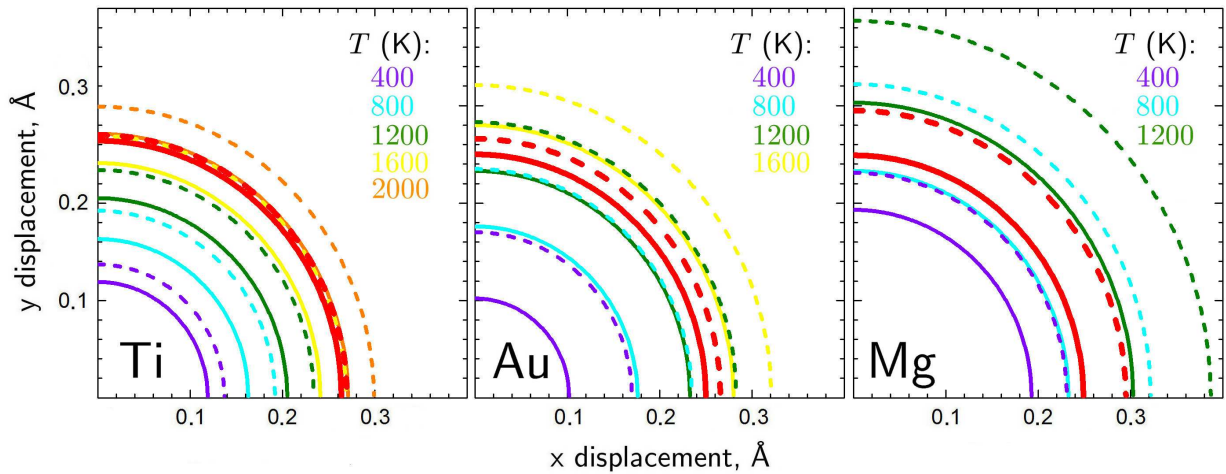}
\caption{
Potential energy surface for 6 nm-radius metal nanoparticles whose constituent
atoms interact via the original (dashed lines) or the modified (solid lines) potentials.
The thick dashed and solid (red) lines denote the energy difference corresponding to
the predicted bulk melting temperatures (see Table~\ref{Table_Tmelt}).
}
\label{fg:potential}
\end{figure*}

Our analysis has revealed that interatomic interactions at distances, exceeding the
equilibrium distance by a characteristic vibration amplitude defined by the Lindemann
criterion, significantly affect the correctness of simulations.
To elaborate on this issue, the following procedure has been adopted.
We have analyzed the potential energy surfaces (PES) for the studied metal systems.
As a case study, we considered large 6 nm-radius nanoparticles with the optimized
structure; positions of all atoms in the system except for a given one were fixed.
The movable atom was displaced from its equilibrium position and the interaction energy
was calculated.
Then, the energy of the perturbed system was subtracted from the energy of the fully
optimized system.
The resulting PES for the metal nanoparticles are presented in Figure~\ref{fg:potential}.
Each panel shows several isolines corresponding to a given energy difference between
the optimized and the perturbed systems.
For the sake of clarity, this quantity has been converted into temperature.

The figure illustrates that the modified potential (solid curves),
due to the additional linear term, makes the resulting potential steeper at
large interatomic distances, as compared to the original EAM-type potential (dashed curves).
For instance, in the case of titanium (left panel), the displacement of
an atom for about 0.3~\AA, that is approximately 1/10 of the closest interatomic
distance ($d_{\rm Ti} = 2.95~\AA$), results in the energy difference of
about 0.17~eV that corresponds to 2000~K.
Thus, interatomic interactions at distances, exceeding the equilibrium distance
by a characteristic vibration amplitude $\delta u$, are overestimated by
conventional EAM-type potentials and should be corrected in order to reproduce
the quantitatively correct value of the melting point.
A more accurate description of the interatomic interaction in the region beyond
the equilibrium distance allows one to handle the problem of the accurate description
of thermal properties of metal materials.

\section{Conclusion}

In summary,
we have formulated a recipe for modifying the embedded-atom method-type potential
that reconciles the simulated melting temperature and ground-state properties
of metals by means of molecular dynamics simulations.
It has been demonstrated that the modified many-body potential reduces the gap
between the simulated and the experimental values of bulk melting temperature of
metal systems, such as titanium, magnesium, gold and platinum, down to about a few
percent and also does not affect the accuracy of description of ground-state
properties like the lattice parameters, cohesive energy
and the energy of vacancy formation.
The physical background behind this improvement is that the modified potential
weakens the interatomic interactions at large distances, which are typically
overestimated by the conventional embedded-atom method.
A proper account for the long-distance interatomic interactions has been found
to be crucial for a quantitatively accurate simulation of melting and other
excited vibrational state properties of the system being far from the potential
energy minimum.
The introduced modification of the embedded-atom method-type potential can be
utilized for an accurate modeling of other metals and alloys that have been
treated with the use of this approach so far.

%%%%%%%%%%%%%%%%%%%%%%%%%%%%%%%%%%%%%%%%%%%%%%%%%%%%%%%%%%%%%%%%%%%%%
\section*{Acknowledgements}

The work was supported by the European Commission (the FP7 Multi-ITN Project
''ARGENT'', grant agreement no.~608163).
The possibility to perform computer simulations at the Frankfurt Center
for Scientific Computing using the LOEWE-CSC cluster is gratefully acknowledged.
AVK acknowledges the support from the Alexander von Humboldt Foundation.

%%%%%%%%%%%%%%%%%%%%%%%%%%%%%%%%%%%%%%%%%%%%%%%%%%%%%%%%%%%%%%%%%%%%%
%\section*{References}

\end{document}